\documentclass[nofootinbib,twocolumn,showpacs,preprintnumbers,
amsmath2000,amssymb]{revtex4}
\usepackage{graphicx}% Include figure files
\usepackage{dcolumn}% Align table columns on decimal point
\usepackage{bm}% bold math

\begin{document}

%%%%%% private definition %%%%%%%
\def\CR{\nonumber \\}
\def\pt{\partial}
\def\be{\begin{equation}}
\def\ee{\end{equation}}
\def\bea{\begin{eqnarray}}
\def\eea{\end{eqnarray}}
\def\eq#1{(\ref{#1})}
\def\la{\langle}
\def\ra{\rangle}
\def\hyp{\hbox{-}}
%%%%%%%%%%%%%%%%%%%%%%%%%%%%%%%%%

\preprint{YITP-02-13, hep-th/0203032}

\title{Analytic continuations of de Sitter thick domain wall solutions}
\author{Naoki Sasakura}
\email{sasakura@yukawa.kyoto-u.ac.jp}
\affiliation{Yukawa Institute for Theoretical Physics, Kyoto University,
Kyoto 606-8502, Japan}

\date{\today}

\begin{abstract}
We perform some analytic continuations of the de Sitter thick domain
wall solutions obtained in our previous paper \cite{Sasakura} in the system 
of gravity and a scalar field with an axion-like potential. 
The obtained new solutions represent anti-de Sitter thick domain walls 
and cosmology. The anti-de Sitter domain wall solutions are periodic, 
and correspondingly the cosmological solutions represent cyclic universes. 
We parameterize the axion-like scalar field potential and determine 
the parameter regions of each type of solutions.
\end{abstract}


\maketitle

\section{Introduction}

The solitonic objects in string theory play major roles in the understanding
of its non-perturbative properties such as dualities.
In string theory the solitons couples necessarily with gravity. 
In view of the universality of the gravitational interaction, 
the understanding of the gravitational
aspects of solitons may provide deeper understanding of the variety of 
the dynamics contained in string theory.

The simplest system of solitons and gravity would be domain walls
of scalar field theory interacting with gravity.
In field theory, domain walls appear when there are more than one
vacua in a scalar field potential, and it is a straightforward
matter to obtain static domain wall configurations. 
On the other hand, with gravitational interactions,  
their non-linearity and instability make it a non-trivial issue to 
study the dynamics of domain walls. 
As for a supersymmetric domain wall configuration, it can be analyzed 
by BPS first order differential equations \cite{Cvetic}.
These solutions are static and it was shown that a generic static and flat
domain wall configuration can be analyzed in the same manner 
\cite{Csaki,Skenderis,Chamblin,DeWolfe}. 
The procedure to obtain such a solution is mathematically well-defined, 
but an obtained solution is not necessarily physically meaningful.
In fact, the scalar field potential must be fine-tuned to avoid naked 
curvature singularities \cite{DeWolfe,Gremm:2000dj,Flanagan:2001dy,Sasakura}. 
This situation is quite unsatisfactory, since a scalar field potential 
will change its form from a supersymmetric one by possible low energy
dynamics such as supersymmetry breaking and instanton corrections,
even if we assume some high energy supersymmetries.
The non-supersymmetric string theories
also suffer from a similar pathological behavior 
that the background solutions 
have naked singularities where the dilaton field diverges
\cite{Dudas:2000ff,Blumenhagen:2001dc}.

One of the possible resolutions to these singularities is to 
introduce time-dependences of domain walls.
The use of a de-Sitter expansion to turn a curvature singularity
of a static soliton to a horizon has appeared in the context of global
$U(1)$ vortex solutions \cite{Gregory:1996dd} and 
in a codimension two non-supersymmetric soliton solution in IIB string 
theory \cite{Berglund:2001aj}. This is also used in constructing 
background solutions of non-supersymmetric string 
theories \cite{Charmousis:2002nq}.
As for time-dependent domain walls, perturbative analyses have been performed
for the system of gravity and a scalar field
in \cite{Bonjour:1999kz,Bonjour:2000ca}.
The constructions of analytic de Sitter thick domain wall solutions with 
horizons were done in  \cite{Wang:2002pk,Sasakura}.

In our previous paper \cite{Sasakura}, the scalar field potential
takes an axion-like form and its two parameters are restricted to
a certain region for the analytic solutions to exist. 
The motivation of this paper is to find analytical
solutions for the outside of this region of these parameters.
It is a well known trick that, starting from a domain wall solution,
analytic continuations generate domain wall solutions 
with flipped curvatures and cosmological solutions \cite{Cvetic:1997vr}.
We will show that the new solutions cover the missing parameter regions.
They are anti-de Sitter domain walls, finite lifetime universes with a 
big-bang and a big-crunch and cyclic universes.

\section{De Sitter domain wall solutions}

In our previous paper \cite{Sasakura}, 
we have obtained a class of analytic solutions of thick
domain walls with de Sitter expansions in the system of five-dimensional 
gravity and a scalar field with an axion-like potential.
Let us start our discussions by extending our previous results to
a general space-time dimension $n$.

The action of our system is given by 
\be
S=\int dt dx^{n-2} dy \ \sqrt{-g} \left(R - \frac12 g^{\mu\nu} 
\pt_\mu \phi \pt_\nu \phi-V(\phi)\right).
\ee
The metric ansatz we use is the warped geometry
\be
\label{warplor}
ds^2=a(y)^2 \left( -dt^2+e^{2Ht}\sum_{i=1}^{n-2}(dx^i)^2\right) +dy^2,
\ee
where $H$ denotes the Hubble constant of the $(n-1)$-dimensional de Sitter
space-time.
Under the assumption that the scalar field depends only on the coordinate $y$,
the Einstein equations are
\bea
\label{eeq1}
(\phi')^2&=& \frac{2(n-2)\left((a')^2-a a''-H^2\right)}{a^2},\cr
V(\phi)&=&\frac{(n-2)\left(-a a''-(n-2) (a')^2+(n-2)H^2\right)}{a^2}, 
\eea
where $'$ denotes the derivation with respect to $y$.
The equation of motion of the scalar field is automatically satisfied by 
the solutions of \eq{eeq1} because of a Bianchi identity.
By repeating the same procedure done in our previous paper \cite{Sasakura}, 
we find that the first equation of \eq{eeq1} is satisfied by the solution
\bea
\label{solution}
a(y)&=&{\rm sn}(H y,i\beta^{-1}),\cr
\phi(y)&=&\sqrt{2(n-2)} \arctan
\left(\frac{{\rm cn}(H y,i\beta^{-1})}{\beta {\rm dn}(H y,i\beta^{-1})}\right),
\eea
where the elliptic functions are defined by 
\be
{\rm sn}^{-1}(z,k)=\int_0^z\frac{dx}{\sqrt{(1-x^2)(1-k^2x^2)}},
\ee   
${\rm cn}(u,k)=(1-{\rm sn}^2(u,k))^{1/2}$ and 
${\rm dn}(u,k)=(1-k^2{\rm sn}^2(u,k))^{1/2}$.
Here the parameter $\beta$ is a free real parameter.
In the expression \eq{solution}, the normalization of the scale factor $a(y)$ 
merely defines the unit of length scale 
and we have normalized it by imposing $a=1$ at the domain wall peak 
defined by $a'=0$,
and the constant shift ambiguity of $\phi(y)$ is fixed by imposing 
$\phi=0$ at the domain wall peak.
A peculiar property of the solution \eq{eeq1} is that the scale factor
behaves near its vanishing point $y=0$ as 
\be
\label{regular}
a(y)= Hy +O(y^3).
\ee
As discussed in our previous paper \cite{Sasakura}, this behavior
is required for the vanishing point to be regular. 
The factor $H$ of the linear term can be also understood 
from the physical consistency
of the geometry that the temperature associated to the Rindler space-time 
near $y=0$ should agree with that of the de Sitter domain wall 
space-time $\frac{H}{2\pi}$.
We will show in section \ref{extension} 
that the vanishing point is actually a horizon and 
that a regular extended space-time can be 
obtained by taking an appropriate coordinate system.

The scalar field potential is determined by the second equation of
\eq{eeq1}, and we obtain
\bea
\label{potential}
V(\phi)&=&\frac{H^2 (n-2)^2 (1-\beta^{-2})}{2} \cr
&&+\frac{H^2 n(n-2) (1+\beta^{-2})}{2}\cos\left( \sqrt\frac{2}{n-2} 
\phi\right),
\eea
which is similar to that of an axion with an instanton correction.
In this paper we consider the two-dimensional parameter space $(v_0,v_1)$
of the scalar field potential,
\be
\label{parametrize}
V(\phi)=v_0+v_1 \cos\left( \sqrt\frac{2}{n-2} \phi\right).
\ee
We may assume $v_1$ to be a positive parameter by shifting $\phi$,
appropriately.
In this parameterization the result \eq{potential} shows that the parameter 
region for the existence of a regular de Sitter domain wall solution is 
given by
\be
\label{dsregion}
-(n-2) v_1 < n v_0 < (n-2) v_1.
\ee
\begin{figure}[htdp]
\includegraphics{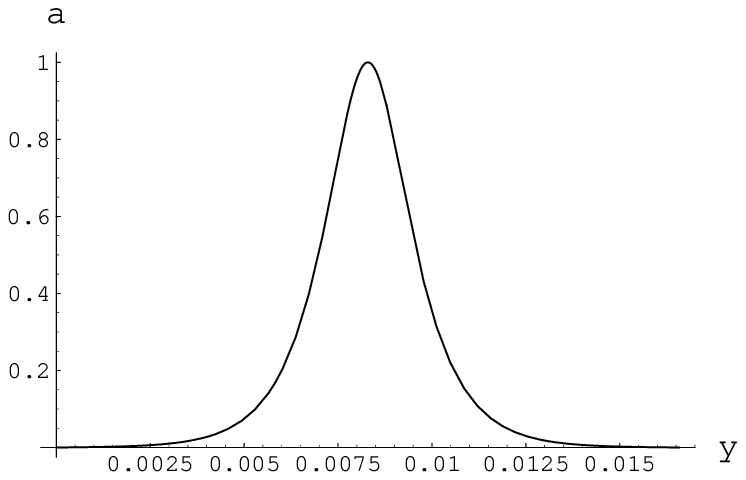}
\includegraphics{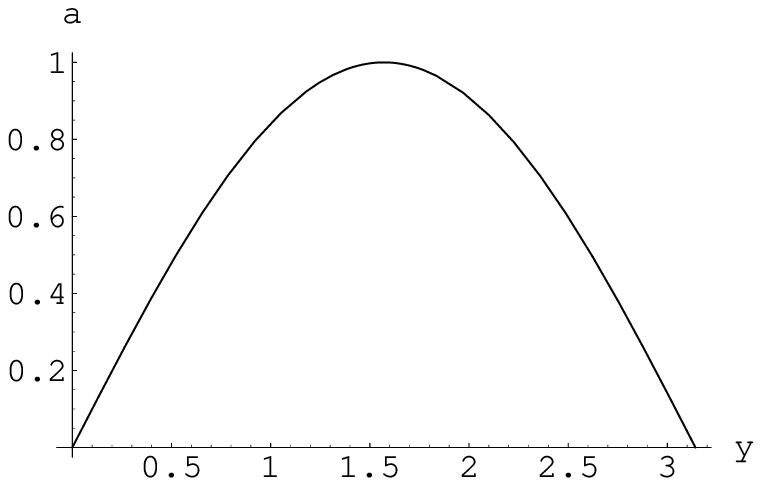}
\caption{\label{figds}The shapes of the de Sitter domain walls with $H=1$ and 
$\beta=0.001$ and $\beta=1000$.}
\end{figure}

\section{\label{extension} Extension of the de Sitter domain wall 
space-time}

The vanishing points $a=0$ are actually horizons. To show this we will 
extend the solutions \eq{solution}.
The extended space-time is essentially equivalent to
the $n=1/2$ case of \cite{Wang:2002pk}, and we will follow their discussions.
We first change to a conformal coordinate $dz=dy/a$.
Integrating the solution \eq{solution}, we obtain
\be
\label{z}
z-z_0=\frac1H \ln\left( \frac{{\rm sn}(Hy,i \beta^{-1})}{
{\rm dn}(Hy,i \beta^{-1})+{\rm cn}(Hy,i \beta^{-1}) }\right),
\ee
where $z_0$ is an integration constant.
Using \eq{z} and taking a proper value of $z_0$, the metric for a de Sitter 
domain wall solution becomes 
\bea
ds^2&=&\frac{2}{(1+\beta^{-2})\cosh (2 H z)+1-\beta^{-2}} \cr
&&\times \left(
-d\tau^2+dz^2+H^{-2}\cosh^2(H\tau)d\Omega^{n-2}\right),
\eea
where we have changed to a global coordinate of a de Sitter space-time instead
of a flat one appearing in \eq{warplor}, and $d\Omega^{n-2}$ denotes the 
metric on an $(n-2)$-dimensional unit sphere.
By a further transformation of $u=\exp(H(\tau-z))$ and $v=\exp(-H(\tau+z))$, 
we obtain
\bea
ds^2=\frac{4H^{-2}}{(1+\beta^{-2})(1+u^2v^2)+2(1-\beta^{-2})uv} \cr
\times \left(du dv+\left(\frac{u+v}{2}\right)^2 d\Omega^{n-2}\right).
\label{uvmetric}
\eea
Note that the allover factor of the metric \eq{uvmetric} is non-singular
for all the real values of $u$ and $v$.
After the change of variables $u=T+R,\ v=-T+R$, 
the metric in the parentheses becomes 
$-dT^2+dR^2+R^2 d\Omega^{n-2}$ and is free of a conical singularity.
As pointed out in \cite{Wang:2002pk}, 
the domain wall peak at $-T^2+R^2=uv=1$ is 
a bubble with a constant acceleration in the coordinate $(T,R,\Omega)$.

As for the scalar field, we obtain
\be
\phi=\sqrt{2(n-2)}\arctan\left(\frac{uv-1}{\beta(uv+1)}\right),
\ee
which is also well-defined for any real values of $u$ and $v$.
\begin{figure}[htdp]
\includegraphics{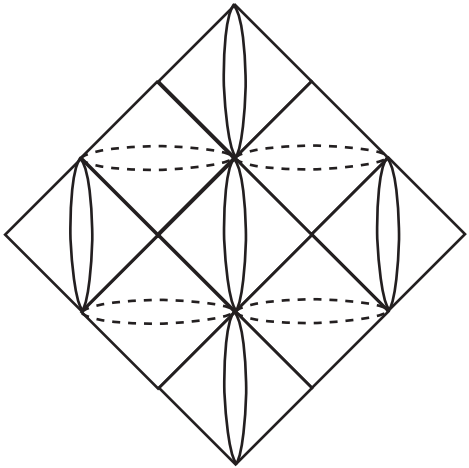}
\caption{\label{penrose} 
The Penrose diagram of the de Sitter domain wall space-time.
The two edges of each oval should be identified. The solid ones depict 
the peaks of the domain walls, while the dashed ones the bottoms.}
\end{figure}

\section{Anti-de Sitter domain wall solutions}

In the following we will study the other parameter regions of the 
potential \eq{parametrize} than \eq{dsregion}.
Looking at the expression \eq{potential}, one notices that 
all the other parameter regions of 
$v_0$ and $v_1$ can be covered by the analytic continuations of the 
parameters $\beta$ and $H$ to pure imaginary values.
However, under an analytic continuation of only one of $\beta$ or $H$ to a 
pure imaginary value, one of $\phi(y)$ or $a(y)$ becomes imaginary 
and physically meaningless. 
This cannot be resolved even by using the constant shift ambiguity 
of $y$ of the solution.
To obtain a physically meaningful expression from \eq{solution}, 
both of $\beta$ and $H$ must be analytically continued to imaginary values,
and, after some trials, it turns out that the analytic continuation should
contain a simultaneous shift of $y$ as 
\footnote{The other available
choices of the shift of $y$ do not give any other independent solutions.} 
\bea
\label{continuation}
1/\beta &\rightarrow& -i \delta, \cr
H&\rightarrow& i h, \cr
Hy &\rightarrow& ihy+K(\delta),
\eea
where $\delta$ and $h$ are real constants and $K(\delta)$ is the elliptic 
integral of the first kind,
\be
K(\delta)=\int_0^1\frac{dx}{\sqrt{(1-x^2)(1-\delta^2x^2)}}.
\ee
Substituting the analytic continuation \eq{continuation} into the 
de Sitter solution \eq{solution}, we obtain new solutions 
\bea
\label{adssolution}
a(y)&=&\frac{1}{{\rm dn}(hy,\sqrt{1-\delta^2})},\cr
\phi(y)&=& \sqrt{2(n-2)}\arctan\left(
\frac{\delta~{\rm sn}(hy,\sqrt{1-\delta^2})}{{\rm cn}(hy,\sqrt{1-\delta^2})} 
\right), 
\eea
and
\bea
\label{adspotential}
V(\phi)&=&-\frac{h^2(n-2)^2(1+\delta^2)}{2} \cr
&&-\frac{h^2 n(n-2) (1-\delta^2)}{2}
\cos\left( \sqrt\frac{2}{n-2} \phi\right).
\eea
By the parameterization \eq{parametrize}, the scalar potential 
\eq{adspotential} is in the region
\be
\label{adsregion}
n v_0 < -(n-2) v_1.
\ee
\begin{figure}[htdp]
\includegraphics{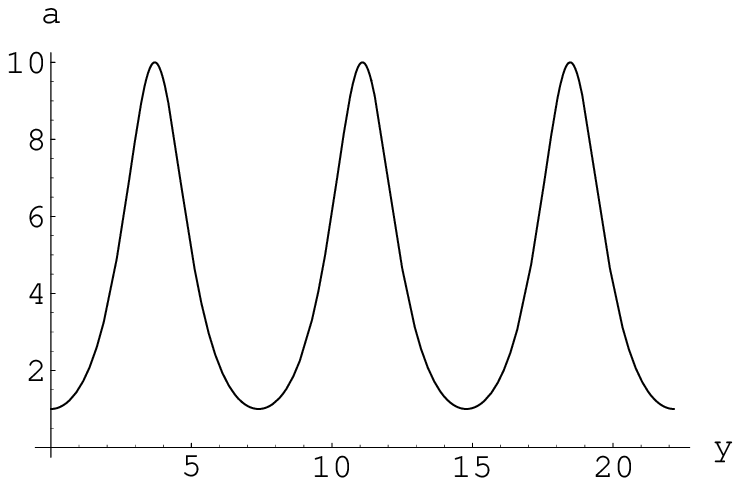}
\caption{\label{ads01}The shape of the anti-de Sitter domain wall 
with $h=1$ and $\delta=0.1$.}
\end{figure}

To make the warped metric \eq{warplor} meaningful under the analytic 
continuation \eq{continuation}, we perform a simultaneous continuation 
$t\rightarrow i r$ and $x^1\rightarrow ix^0$. 
Then \eq{warplor} becomes an anti-de Sitter domain wall metric
\be
\label{adsmetric}
ds^2=a(y)^2 \left(dr^2+e^{-2hr}\left(-(dx^0)^2+\sum_{i=2}^{n-2} (dx^i)^2\right)
\right)+dy^2.
\ee  
Thus in the region \eq{adsregion}, there exist regular AdS domain wall
solutions, which are given by \eq{adssolution} and \eq{adspotential}. 

The stability of the AdS solutions can be checked as follows. 
We restrict our attention to the five dimensional case ($n=5$).
Presumably the extension to a general dimension will be straightforward. 
As discussed in \cite{DeWolfe,Gremm:2000dj,DeWolfe:2000xi,Kobayashi:2001jd},
the problem of obtaining the mass spectra of the linear perturbations
around a solution boils down to solving Schrodinger equations.
As for the tensor perturbation,
\bea
ds^2&=&a^2((\gamma_{\mu\nu}+h_{\mu\nu})dx^\mu dx^\nu+dz^2),\cr
{h_{\mu}}^\mu&=&{h_{\mu\nu}}^{|\nu}=0,
\eea
the Schrodinger equation has a supersymmetric form
\be
\label{qqeq}
Q^\dagger Q\varphi=m^2\varphi,
\ee
where 
\be
Q=\frac{d}{dz}-\frac3{2a}\frac{da}{dz}.
\ee
Thus the eigenvalues $m^2$ is non-negative, and hence  
the stability under the tensor perturbations is satisfied.
As for the scalar perturbation,
\be
ds^2=a^2((1+\psi_1)\gamma_{\mu\nu}dx^\mu dx^\nu+(1+\psi_2)dz^2),
\ee
the Schrodinger equation 
becomes \cite{DeWolfe:2000xi,Kobayashi:2001jd,Sasakura}
\be
\label{schscal}
\left(\tilde{Q}^\dagger \tilde{Q} + 
\frac{a^2(\phi')^2}3+4 h^2 \right)\varphi=m^2\varphi,
\ee
where 
\be
\tilde{Q}=\frac{d}{dz}+\frac {d \ln (a^{3/2}\phi')}{dz}.
\label{defqtilde}
\ee
This Schrodinger equation is obtained by the substitution 
$H\rightarrow ih$ in the corresponding expression in \cite{Sasakura}.  
The $m^2$ is obviously positive,
and the AdS solutions are stable under the scalar perturbations.
However there remains one thing to check before this conclusion.
If there existed points with $\phi'=0$,
the operator $\tilde{Q}$ would become singular at these points and  
the above naive discussion of the positivity would be in danger.
Moreover, in the derivation of the Schrodinger equation for the scalar 
perturbations in \cite{Kobayashi:2001jd}, the linear perturbation is 
redefined by a multiplication of a factor which is singular if $\phi'=0$. 
But, in the solution \eq{adssolution}, $\phi'$ is always non-zero,
and the above discussion is safe.

\section{ The region without regular solutions}
We cannot find a physically meaningful analytic continuation to the parameter
region 
\be
(n-2) v_1<n v_0.
\label{noregion}
\ee
Thus it is suspected that there are no regular 
domain wall solutions of the form of the warped metric \eq{warplor}
in this parameter region. 
To see whether this is the case, we will use a numerical computation.

Since, in this parameter region, the constant part $v_0$ 
of the scalar potential 
is larger than that of the parameter region of de Sitter analytic 
solutions \eq{dsregion}, 
we may assume the solution to be a de Sitter domain wall rather than 
an anti-de Sitter one. 
As generally discussed in our previous paper \cite{Sasakura}, 
a regular de Sitter domain wall solution is sandwiched between two horizons.
The initial values of the differential equations \eq{eeq1} can be provided by
the three values $a$, $a'$ and $\phi$ at a certain initial location of $y$. 
For the solution to be regular at a horizon, the scale factor must behave
in the form \eq{regular}, and therefore the freedom to choose the initial 
values is reduced to the only free parameter $\phi=\phi_c$ at the horizon.
Thus, to search for a regular solution, 
we take one of the horizons as the initial location, and   
integrate numerically the differential equations \eq{eeq1} for each value of
$\phi_c$ at the horizon.
Then the question of the existence of a regular solution is translated to
whether there
exists an initial value $\phi_c$ for which the numerical solution of the 
differential equations is regular between the initial horizon and the other
horizon of a domain wall.

This search procedure gives a numerical regular solution for each choice of 
$(v_0,v_1,H)$. Studying in this three-dimensional space would be too much, 
and in fact, we can reduce the dimension to one 
by rescaling the differential equations \eq{eeq1}. 
By the rescaling of $y$ and $a$, we can normalize the scalar potential 
and the Hubble constant so that $v_1=1$ and $H=1$.
Thus it is enough to check the question for each choice of $v_0$.

It would be a reasonable assumption that a domain wall contains 
the peak of the potential energy $\phi=0$. Then, since the potential 
\eq{parametrize} is $Z_2$ invariant\footnote{Namely, invariant 
under $\phi\rightarrow -\phi$.}, 
it is enough for us to solve the equations in the region $\phi>0$.
We take an initial value $\phi=\phi_c>0$ and solve the differential
equations \eq{eeq1} taking the branch $\phi'<0$, until the solution reaches 
the value $\phi=0$.
By sweeping the initial value $\phi_c$, we obtain the range of $a'/a$ 
at $\phi=0$. 
If the obtained range of $a'/a|_{\phi=0}$ contains both positive and negative
values, one can construct a regular solution by gluing a numerical solution 
with a certain value of $p=a'/a|_{\phi=0}$ 
to the $Z_2$ image of the solution with $a'/a|_{\phi=0}=-p$.
If the range does not contain both signs, one cannot construct a 
regular solution. 

\begin{figure}[htdp]
\includegraphics{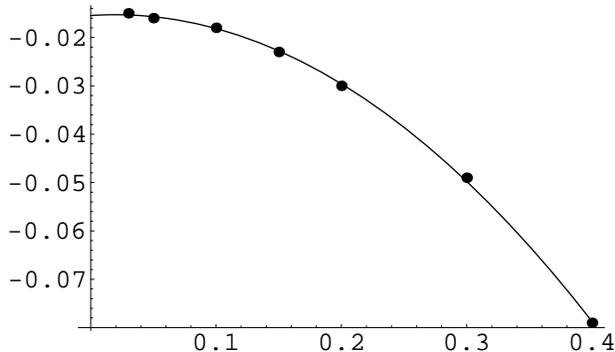}
\caption{\label{noregular} 
The values of $a'/a|_{\phi=0}$ are plotted against $\phi_c$ for
$v_0=11/15,\ v_1=1,\ H=1$.
The dots are the points where the numerical computations were performed, 
and the solid line is the fitting line with an assumed form 
$a'/a|_{\phi=0}=c_0+c_1 \phi_c + c_2 {\phi_c}^2$.}
\end{figure}

We performed the above procedure in five space-time dimensions $(n=5)$. 
From \eq{dsregion}, the maximum value for the existence of an analytic 
solution is $v_0=3/5=9/15$. In fact, for $v_0=7/15$, we obtained 
$a'/a|_{\phi=0}\approx 0.0023,\ -0.015$ for $\phi_c=0.2 \sqrt{3/2}\pi,\ 
0.3 \sqrt{3/2}\pi$, respectively, and the existence of a
regular solution is numerically supported.
For $v_0=11/15$, however, the plotted values of $a'/a|_{\phi=0}$ in 
fig.\ref{noregular} 
indicate that $a'/a|_{\phi=0}$ takes only negative values. 
Thus it is numerically supported that there do not exist any regular de 
Sitter domain wall solutions in the parameter region \eq{noregion}.

Qualitatively, in this parameter region, the constant part of the scalar field 
potential is so large that the expansion rate at the core is too large for 
a domain wall to keep its shape. Similar phenomena are discussed in the
context of topological inflation in \cite{Vilenkin:1994pv,Linde:1994hy,
Bonjour:1999kz}. 
In \cite{Bonjour:1999kz}, a perturbative analysis of the phase boundary 
between the existence and non-existence
of domain wall solutions for the four-dimensional system of gravity and 
a scalar field was performed, 
including the present case with an axion-like scalar field
potential\footnote{Denoted as a sine-Gordon potential in 
\cite{Bonjour:1999kz}.}. According to the paper, the
phase boundary is generally characterized by the equation
\be
\frac32 |V''(0)|-V(0)=0
\ee
in our present notation for $n=4$.
Substituting the parameterization \eq{parametrize}, 
this becomes $2v_0-v_1=0$, which agrees with \eq{noregion} for $n=4$. 

Thus, in the region \eq{noregion},
the rapid expansion will ultimately sweep away the spatial 
dependence of the scalar field, and the dynamics will be
mainly described by its time-dependence. 
This will be the subject of section \ref{cosmology}.

\begin{figure}
\includegraphics{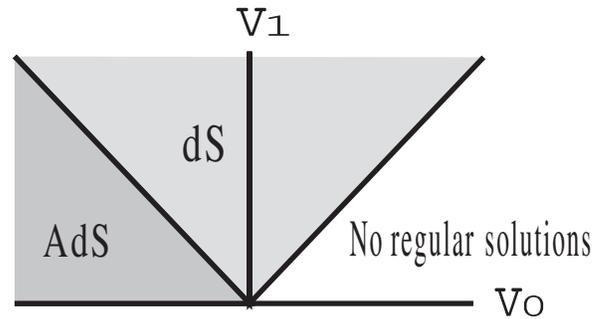}
\caption{\label{wallregion}
The parameter regions for dS, AdS and no regular domain wall solutions
are shown. The two lines of boundaries are $n v_0+(n-2)v_1=nv_0-(n-2)v_1=0$.
On the line $n v_0+(n-2)v_1=0$, flat domain wall solutions exist. }
\end{figure}

\section{\label{cosmology} Cosmological solutions}

It is well known that, starting from a domain wall solution,
a cosmological solution can be obtained by an analytic continuation which 
exchanges the transverse coordinate and the time 
coordinate \cite{Cvetic:1997vr}.  
An appropriate analytic continuation is given by
\bea
\label{concos}
t&\rightarrow&iy, \cr
y&\rightarrow&it, \cr
H&\rightarrow&ih,
\eea
by which, a de Sitter domain wall solution \eq{solution}
changes to a new solution with the substitution $y\rightarrow t$ and 
$H\rightarrow h$. 
This analytic continuation turns the de Sitter domain wall 
metric \eq{warplor} into
\be
\label{metriccos}
ds^2=-dt^2+a(t)^2 \left(dy^2+e^{-2hy}\sum_{i=1}^{n-2}(dx^i)^2\right),
\ee
which describes a FRW cosmology of an open universe.
Hence the new solution represents a finite lifetime open universe 
with a big-bang and a big-crunch.

Under the analytic continuation \eq{concos}, the equations of motion 
\eq{eeq1} remain the same with the identification of $'$ with 
the time derivative 
$\dot{}$ and $H$ with $h$ and changing the sign of the scalar potential.
Under the change of the sign of the potential, the region of the 
de Sitter solutions \eq{dsregion} is transformed to the identical region. 
Thus the parameter region for the open universe solution is the same 
as that of the de Sitter wall solutions \eq{dsregion}.

\begin{figure}[htdp]
\includegraphics{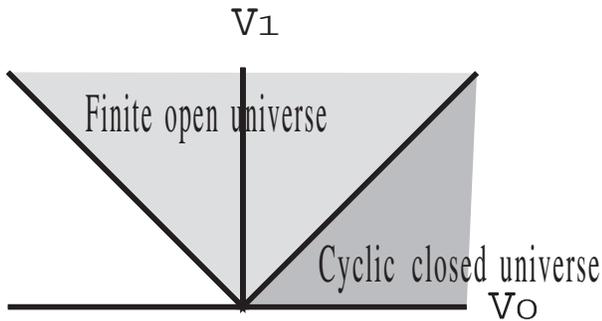}
\caption{\label{uniregion}
The parameter regions for the two types of cosmological solutions
are shown. The two lines of boundaries are $n v_0+(n-2)v_1=nv_0-(n-2)v_1=0$.}
\end{figure}

As for the anti-de Sitter solution, the analytic continuation \eq{concos}
turns it into a cosmological solution of a closed universe.
Because of the flip of the sign of the scalar potential, the parameter region 
of the closed universe solution is obtained by flipping \eq{adsregion} into 
\be
(n-2)v_1<nv_0.
\ee
This agrees with the region where there are no regular domain wall
solutions. 
The corresponding cosmological solutions are obtained by 
the substitution $y\rightarrow t$ and $H\rightarrow h$
in the AdS domain wall solutions \eq{adssolution}.
The periodic behavior of the AdS domain wall solutions is 
now interpreted as representing a cyclic universe.

\begin{figure}[htdp]
\includegraphics{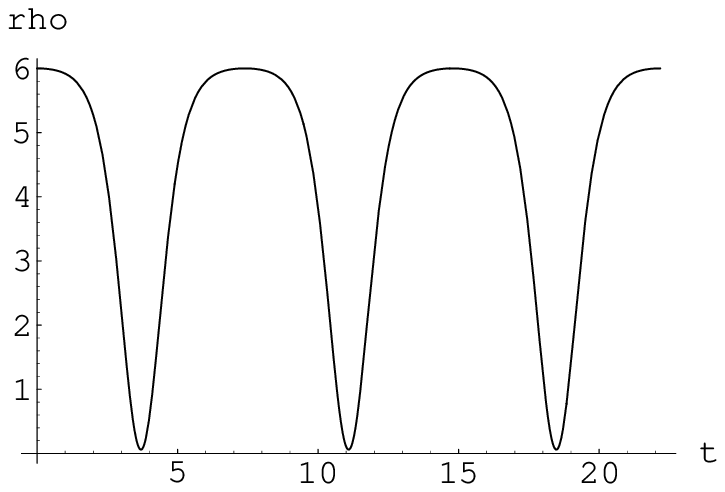}
\includegraphics{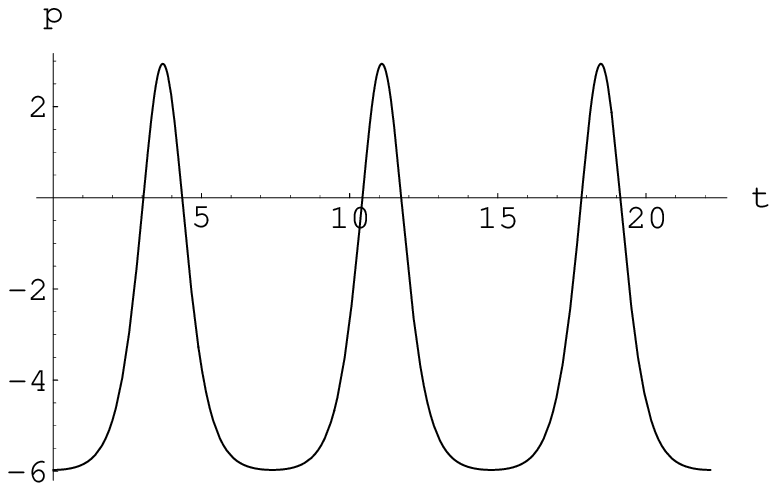}
\caption{\label{figenergy}
The time evolution of the energy density and the pressure of 
the cosmological solution (five dimensions)
corresponding to the AdS solution of fig.\ref{ads01}.
The energy density takes its maximum at the minimum of the scale
factor to initiate the bounce, while the pressure takes its maximum 
at the maximum of the scale factor where the contraction begins. }
\end{figure}

It would be interesting to see the behavior of the energy density 
and the pressure. In a five-dimensional space-time, they are
\bea
\rho&=&\frac14(\dot\phi)^2+\frac12 V(\phi) \cr
&=&6h^2\left(1-
\frac{\delta^2(1-\delta^2){\rm sn}^2(ht,\sqrt{1-\delta^2})}
{{\rm dn}^2(ht,\sqrt{1-\delta^2})}\right), \cr
p&=&\frac14(\dot\phi)^2-\frac12 V(\phi) \cr
&=&h^2 \delta^2
\left(-6-\frac6{\delta^2}+\frac9{{\rm dn}^2(ht,\sqrt{1-\delta^2})}\right).
\eea
In fig.\ref{figenergy}, we plotted the time evolution of the energy density
and the pressure for the cyclic universe solution with $h=1$ and $\delta=0.1$.
At the minimum of the scale factor the energy density takes its maximum 
value and the pressure is approximately the minus of 
the energy density. At this period the large energy density initiates a bounce
and the universe is at the inflation stage.
On the other hand, at the maximum of the scale factor, the energy 
density is nearly vanishing and the pressure dominates. 
The large pressure works as a negative gravitational energy 
and initiates a contraction of the universe.

For a flat FRW universe, if we assume the null energy condition, the Hubble 
parameter satisfies an inequality
\be
\dot H\leq 0,
\label{hubble}
\ee 
and there cannot exist a bounce. This also implies that, since our universe is 
expanding at present days, the universe must have started from a singularity.
In the papers \cite{Khoury:2001bz,Seiberg:2002hr}, 
some loopholes of the above discussion are presented from
string theory to support the possibility of the pre-big bang 
scenario \cite{Veneziano:2000pz}.  
In our cyclic solution, the existence of the positive curvature of 
the space changes the equation of motion of $H$ into     
\be
\dot H=\frac{h^2}{a^2}-\frac{\dot\phi^2}{2(n-2)},
\ee
and $\dot H$ can take both signs.
Considering the high energy at the big-bang of the universe, it is plausible
that some fluctuations of matters and gravity generate a region
with a positive spatial curvature and the bounce of the universe  
simply happens from a classical dynamics.
In this case, even though the present model seems far from what is the real 
universe, it might provide a simple toy model to study 
the pre-big bang scenario.

The inequality \eq{hubble} plays also an important role
in the dS/CFT correspondence. The central charge is given by an 
inverse power of the Hubble parameter, and the inequality \eq{hubble} 
supports the interpretation of the time evolution as a renormalization 
group flow to UV \cite{Strominger:2001gp,Balasubramanian:2001nb}. 
In the case of a cyclic universe, however, the time flow cannot be
interpreted in this way \cite{Argurio}, because reverse process exists. 
If we take seriously the interpretation of 
\cite{Strominger:2001gp,Balasubramanian:2001nb}, there should be 
a kind of mechanism to prevent the above situation to happen.
{}From this point, it would be interesting to try to  
consistently embed our simple model into string theory.     

\section{\label{discussions} Summary and discussions}

In this paper, we have studied the analytic solutions of the system of 
gravity and a scalar field with an axion-like potential.
They contain de Sitter thick domain walls,
anti-de Sitter thick domain walls, finite lifetime universes with a 
big-bang and a big-crunch, and 
cyclic universes. These analytic solutions might be useful as toy 
models for the studies of the more general corresponding cases.    

An obvious application of the analytic de Sitter domain wall solutions 
presented in this and previous papers \cite{Sasakura} 
would be as toy models of our world through the brane world scenario.
According to the recent observations \cite{Fukugita:2000ck}, 
our universe was in the inflation stage at the big-bang, and moreover 
a tiny cosmological constant might exist even at present.
In our model, gravity will be confined by the mechanism of 
\cite{Randall:1999vf}, and it would be interesting to investigate 
the gravitational properties in such an accelerating domain wall universe. 

Another interesting direction would be to embed our model into 
supersymmetric theories or superstring theory.
The potential \eq{parametrize} has a simple form of an axion,
which would be easily generated by field theory or string theory instanton 
corrections.  
According to \cite{Witten:2001kn}, a de Sitter space-time cannot have
any supersymmetries, and hence supersymmetries must be broken on
a de Sitter domain wall.
In the identification of our world with the domain wall, 
it seems challenging to explain 
the large hierarchy between the observed upper
bound of the cosmological constant and the supersymmetry breaking scale.
Based on a similar motivation,
our model may be regarded as a gravity-coupled analogy of the susy-breaking
domain wall solution presented in \cite{Maru:2001gr,Maru:2001sk}.

Our cosmological solutions of cyclic universes might provide toy models
for the scenario of \cite{Steinhardt:2001vw,Felder:2002jk}.

%%%%%%%%%%%%%%%%%%%%%%%%%%%%%%%%%%%%%%%%%%%%%%%%%%%%%%%%%%%%%
\begin{acknowledgments}
The author would like to thank N.~Sakai for discussions.
The author was supported in part by Grant-in-Aid for Scientific Research
(\#12740150), and in part by Priority Area:
``Supersymmetry and Unified Theory of Elementary Particles'' (\#707),
from Ministry of Education, Science, Sports and Culture, Japan.
\end{acknowledgments}


\begin{thebibliography}{99}

%\cite{Sasakura}
\bibitem{Sasakura}
N.~Sasakura,
``A de Sitter thick domain wall solution by elliptic functions,''
JHEP {\bf 0202}, 026 (2002)
[arXiv:hep-th/0201130].
%%CITATION = HEP-TH 0201130;%%

\bibitem{Cvetic}
M.~Cvetic, S.~Griffies and S.~J.~Rey,
``Static domain walls in N=1 supergravity,''
Nucl.\ Phys.\ B {\bf 381}, 301 (1992)
[arXiv:hep-th/9201007].
%%CITATION = HEP-TH 9201007;%%

\bibitem{Csaki}
C.~Csaki, J.~Erlich, T.~J.~Hollowood and Y.~Shirman,
``Universal aspects of gravity localized on thick branes,''
Nucl.\ Phys.\ B {\bf 581}, 309 (2000)
[arXiv:hep-th/0001033].
%%CITATION = HEP-TH 0001033;%%

\bibitem{Skenderis}
K.~Skenderis and P.~K.~Townsend,
``Gravitational stability and renormalization-group flow,''
Phys.\ Lett.\ B {\bf 468}, 46 (1999)
[arXiv:hep-th/9909070].
%%CITATION = HEP-TH 9909070;%%

%\cite{Chamblin}
\bibitem{Chamblin}
A.~Chamblin and G.~W.~Gibbons,
``Nonlinear supergravity on a brane without compactification,''
Phys.\ Rev.\ Lett.\  {\bf 84}, 1090 (2000)
[arXiv:hep-th/9909130].
%%CITATION = HEP-TH 9909130;%%

%\cite{DeWolfe}
\bibitem{DeWolfe}
O.~DeWolfe, D.~Z.~Freedman, S.~S.~Gubser and A.~Karch,
``Modeling the fifth dimension with scalars and gravity,''
Phys.\ Rev.\ D {\bf 62}, 046008 (2000)
[arXiv:hep-th/9909134].
%%CITATION = HEP-TH 9909134;%%

%\cite{Gremm:2000dj}
\bibitem{Gremm:2000dj}
M.~Gremm,
``Thick domain walls and singular spaces,''
Phys.\ Rev.\ D {\bf 62}, 044017 (2000)
[arXiv:hep-th/0002040].
%%CITATION = HEP-TH 0002040;%%

%\cite{Flanagan:2001dy}
\bibitem{Flanagan:2001dy}
E.~E.~Flanagan, S.~H.~Tye and I.~Wasserman,
``Brane world models with bulk scalar fields,''
Phys.\ Lett.\ B {\bf 522}, 155 (2001)
[arXiv:hep-th/0110070].
%%CITATION = HEP-TH 0110070;%%

%\cite{Dudas:2000ff}
\bibitem{Dudas:2000ff}
E.~Dudas and J.~Mourad,
``Brane solutions in strings with broken supersymmetry and dilaton tadpoles,''
Phys.\ Lett.\ B {\bf 486}, 172 (2000)
[arXiv:hep-th/0004165].
%%CITATION = HEP-TH 0004165;%%

%\cite{Blumenhagen:2001dc}
\bibitem{Blumenhagen:2001dc}
R.~Blumenhagen and A.~Font,
``Dilaton tadpoles, warped geometries and large extra dimensions 
for  non-supersymmetric strings,''
Nucl.\ Phys.\ B {\bf 599}, 241 (2001)
[arXiv:hep-th/0011269].
%%CITATION = HEP-TH 0011269;%%

%\cite{Gregory:1996dd}
\bibitem{Gregory:1996dd}
R.~Gregory,
``Non-singular global strings,''
Phys.\ Rev.\ D {\bf 54}, 4955 (1996)
[arXiv:gr-qc/9606002].
%%CITATION = GR-QC 9606002;%%

%\cite{Berglund:2001aj}
\bibitem{Berglund:2001aj}
P.~Berglund, T.~Hubsch and D.~Minic,
``de Sitter spacetimes from warped compactifications of IIB string  theory,''
Phys.\ Lett.\ B {\bf 534}, 147 (2002)
[arXiv:hep-th/0112079].
%%CITATION = HEP-TH 0112079;%%

%\cite{Charmousis:2002nq}
\bibitem{Charmousis:2002nq}
C.~Charmousis,
``Dilaton spacetimes with a Liouville potential,''
Class.\ Quant.\ Grav.\  {\bf 19}, 83 (2002)
[arXiv:hep-th/0107126].
%%CITATION = HEP-TH 0107126;%%

%\cite{Bonjour:1999kz}
\bibitem{Bonjour:1999kz}
F.~Bonjour, C.~Charmousis and R.~Gregory,
``Thick domain wall universes,''
Class.\ Quant.\ Grav.\  {\bf 16}, 2427 (1999)
[arXiv:gr-qc/9902081].
%%CITATION = GR-QC 9902081;%%

%\cite{Bonjour:2000ca}
\bibitem{Bonjour:2000ca}
F.~Bonjour, C.~Charmousis and R.~Gregory,
``The dynamics of curved gravitating walls,''
Phys.\ Rev.\ D {\bf 62}, 083504 (2000)
[arXiv:gr-qc/0002063].
%%CITATION = GR-QC 0002063;%%

%\cite{Wang:2002pk}
\bibitem{Wang:2002pk}
A.~h.~Wang,
``Thick de Sitter brane worlds, 
dynamic black holes and localization of  gravity,''
arXiv:hep-th/0201051.
%%CITATION = HEP-TH 0201051;%%

%\cite{Cvetic:1997vr}
\bibitem{Cvetic:1997vr}
See for example,
M.~Cvetic and H.~H.~Soleng,
``Supergravity domain walls,''
Phys.\ Rept.\  {\bf 282}, 159 (1997)
[arXiv:hep-th/9604090].
%%CITATION = HEP-TH 9604090;%%

%\cite{DeWolfe:2000xi}
\bibitem{DeWolfe:2000xi}
O.~DeWolfe and D.~Z.~Freedman,
``Notes on fluctuations and correlation functions 
in holographic renormalization group flows,''
arXiv:hep-th/0002226.
%%CITATION = HEP-TH 0002226;%%

%\cite{Kobayashi:2001jd}
\bibitem{Kobayashi:2001jd}
S.~Kobayashi, K.~Koyama and J.~Soda,
``Thick brane worlds and their stability,''
Phys.\ Rev.\ D {\bf 65}, 064014 (2002)
[arXiv:hep-th/0107025].
%%CITATION = HEP-TH 0107025;%%

%\cite{Vilenkin:1994pv}
\bibitem{Vilenkin:1994pv}
A.~Vilenkin,
``Topological inflation,''
Phys.\ Rev.\ Lett.\  {\bf 72}, 3137 (1994)
[arXiv:hep-th/9402085].
%%CITATION = HEP-TH 9402085;%%

%\cite{Linde:1994hy}
\bibitem{Linde:1994hy}
A.~D.~Linde,
``Monopoles as big as a universe,''
Phys.\ Lett.\ B {\bf 327}, 208 (1994)
[arXiv:astro-ph/9402031].
%%CITATION = ASTRO-PH 9402031;%%

%\cite{Khoury:2001bz}
\bibitem{Khoury:2001bz}
J.~Khoury, B.~A.~Ovrut, N.~Seiberg, P.~J.~Steinhardt and N.~Turok,
``From big crunch to big bang,''
Phys.\ Rev.\ D {\bf 65}, 086007 (2002)
[arXiv:hep-th/0108187].
%%CITATION = HEP-TH 0108187;%%

%\cite{Seiberg:2002hr}
\bibitem{Seiberg:2002hr}
N.~Seiberg,
``From big crunch to big bang - is it possible?,''
arXiv:hep-th/0201039.
%%CITATION = HEP-TH 0201039;%%

%\cite{Veneziano:2000pz}
\bibitem{Veneziano:2000pz}
G.~Veneziano,
``String cosmology: The pre-big bang scenario,''
arXiv:hep-th/0002094.
%%CITATION = HEP-TH 0002094;%%

%\cite{Strominger:2001gp}
\bibitem{Strominger:2001gp}
A.~Strominger,
``Inflation and the dS/CFT correspondence,''
JHEP {\bf 0111}, 049 (2001)
[arXiv:hep-th/0110087].
%%CITATION = HEP-TH 0110087;%%

%\cite{Balasubramanian:2001nb}
\bibitem{Balasubramanian:2001nb}
V.~Balasubramanian, J.~de Boer and D.~Minic,
``Mass, entropy and holography in asymptotically de Sitter spaces,''
Phys.\ Rev.\ D {\bf 65}, 123508 (2002)
[arXiv:hep-th/0110108].
%%CITATION = HEP-TH 0110108;%%

\bibitem{Argurio}
R.~Argurio,
``Comments on cosmological RG flows,''
arXiv:hep-th/0202183.
%%CITATION = HEP-TH 0202183;%%

%\cite{Fukugita:2000ck}
\bibitem{Fukugita:2000ck}
See for example, M.~Fukugita,
``Cosmology and particle physics,''
arXiv:hep-ph/0012214.
%%CITATION = HEP-PH 0012214;%%

%\cite{Randall:1999vf}
\bibitem{Randall:1999vf}
L.~Randall and R.~Sundrum,
``An alternative to compactification,''
Phys.\ Rev.\ Lett.\  {\bf 83}, 4690 (1999)
[arXiv:hep-th/9906064].
%%CITATION = HEP-TH 9906064;%%

%\cite{Witten:2001kn}
\bibitem{Witten:2001kn}
E.~Witten,
``Quantum gravity in de Sitter space,''
arXiv:hep-th/0106109.
%%CITATION = HEP-TH 0106109;%%

%\cite{Maru:2001gr}
\bibitem{Maru:2001gr}
N.~Maru, N.~Sakai, Y.~Sakamura and R.~Sugisaka,
``SUSY breaking by stable non-BPS walls,''
in {\it C01-07-03.2}
arXiv:hep-th/0109087.
%%CITATION = HEP-TH 0109087;%%

%\cite{Maru:2001sk}
\bibitem{Maru:2001sk}
N.~Maru, N.~Sakai, Y.~Sakamura and R.~Sugisaka,
``SUSY breaking by stable non-BPS configurations,''
arXiv:hep-th/0112244.
%%CITATION = HEP-TH 0112244;%%

%\cite{Steinhardt:2001vw}
\bibitem{Steinhardt:2001vw}
P.~J.~Steinhardt and N.~Turok,
``A cyclic model of the universe,''
arXiv:hep-th/0111030.
%%CITATION = HEP-TH 0111030;%%

%\cite{Felder:2002jk}
\bibitem{Felder:2002jk}
G.~N.~Felder, A.~Frolov, L.~Kofman and A.~Linde,
``Cosmology with negative potentials,''
Phys.\ Rev.\ D {\bf 66}, 023507 (2002)
[arXiv:hep-th/0202017].
%%CITATION = HEP-TH 0202017;%%

\end{thebibliography}
\end{document}